# N-functionalized Ti$_2$C MXene as high-performance adsorbent for strontium ions: a first-principles study

Kaikai Qiu[a], Yujuan Zhang[a]*, Lin Wang[b], Mingyu Wu[a], Jingyuan Jin[a], Weiqun Shi[b]*

[a]School of Materials Science and Engineering, University of Science and Technology Beijing, 100083, Beijing, China

[b]Laboratory of Nuclear Energy Chemistry and Key Laboratory for Biomedical Effects of Nanomaterials and Nanosafety, Institute of High Energy Physics, Chinese Academy of Sciences, 100049, Beijing, China

*Mail addresses: zhangyujuan@ustb.edu.cn (Yujuan Zhang), shiwq@ihep.ac.cn (Weiqun Shi)

**Abstract:** Radionuclides sequestration through adsorption technology has attracted much attention due to its unique characters such as high removal efficiency, low cost, and ease of operation. In this work, the interaction mechanism of N-functionalized Ti$_2$C MXene (Ti$_2$CN$_2$) as a potential adsorbent for the removal of strontium ions is investigated by using first principles method. Our results show that surface N atoms are connected to Sr ions by a robust chemical N-Sr bond, which provides a stronger interaction and greater capacity (1.291g g$^{-1}$) of Sr ion adsorption on Ti$_2$CN$_2$ than Ti$_2$C with other surficial groups O, F and OH. Furthermore, the thermal stability of the adsorption structure of Sr on Ti$_2$CN$_2$ with full layer coverage at room temperature is verified by using ab initio molecular dynamics simulations. Our results are expected to provide a new perspective for the design of MXene materials as adsorbent for radionuclide.

## 1. Introduction

Nuclear energy as a high energy density, low-carbon and clean energy is an effective choice to optimize the energy structure. With the continuous development of nuclear energy, the hazardous release of radionuclides and a major environmental concern related to nuclear power have become the key factors restricting the sustainable development of nuclear energy[1]. For example, Fukushima nuclear power plant accident, resulted in the discharge of large amounts of emitted nuclear wastewater into the ocean, has revived worldwide fears of nuclear contamination[2-4]. It is urgent to find an efficient and economical method to remove nuclear waste.

Strontium (Sr), as a typical heat-release radionuclide, could lead to the decomposition of solidified radioactive waste during solidification, and pose difficulties in the disposal of nuclear waste. On the other hand, once Sr is released into environment, it will cause great harm to the biosphere as Sr ions have high mobility and solubility in water. Because of similar chemical properties between Sr and Ca and similar atomic radii, Sr can replace Ca and accumulate in the human body, resulting in high health risk to human body, and therefore must be isolated from the biosphere[5, 6]. There are many methods proposed to treat $Sr^{2+}$ in aqueous solutions, such as ion exchange[7, 8], extraction[9, 10], electrocoagulation[11], and selective separation with the use of liquid membranes[12]. But above methods are expensive, complex and inefficient, and can cause secondary contamination, which makes them difficult to use on a large scale. At present, the most practical method for treating Sr is adsorption by virtue of its ease of operation, high efficiency and environmental friendliness[13-15]. For achieving better performance, the development of novel adsorbent materials has earned extensive attention.

Two-dimensional (2D) graphene-like MXenes are an emerging material family with high specific areas, enormous active sites, high hydrophilicity and excellent chemical stability[16]. The general formula of MXene is $M_{n+1}X_nT_x$ ($n=1$, 2 or 3), where M represents a transition metal, X is generally nitrogen or carbon, and T represents surface terminated functional groups[17-20]. Due to their unique characteristics, MXenes have been demonstrated as novel adsorbent for radionuclides including $^{232}$Th, $^{235,238}$U, $^{152,154}$Eu, $^{133,140}$Ba and $^{137}$Cs[21-24]. Recent studies have shown that the adsorption properties of MXene are closely related to the surface functional groups. For example, it was shown that excessive OH functional groups are considered unfavorable for metal ion adsorption, while O functional groups exhibit a high ion adsorption capacity[25]. Therefore, the modification of functional groups may be a very effective way to improve the adsorption performance. In particular, the molten salt preparation method currently makes it possible to modify the surface functional group type in a targeted manner[26-28]. Therefore, systematically studying the adsorption behavior of MXenes related to different surface functional groups is a crucial step to optimize the adsorption performance of these materials.

As a member of the titanium carbide MXenes, $Ti_2CT_x$ possesses a surface structure and physicochemical properties extremely similar to $Ti_3C_2T_x$, but in fewer layers of C and Ti stacks. In view of the excellent adsorption properties for Sr ion demonstrated by $Ti_3C_2T_x$ in previous studies, we believe that $Ti_2CT_x$ with fewer layers may exhibit a more outstanding performance. Moreover, recent studies have shown that N-functionalization can greatly enhance the adsorption capacity of MXene for metal ions[29, 30]. In this work, we systematically investigate the adsorption mechanism of the N-functionalization titanium carbide $Ti_2CN_2$ for Sr ion using first principles method. Compared to $Ti_2CT_2$ (T=OH, F and O), our results show that $Ti_2CN_2$ under ideal conditions would have excellent adsorption properties for Sr radionuclide, including strong adsorption interaction and high adsorption capacity. The strong N-Sr bond contributes to the strong interaction between Sr and the $Ti_2CN_2$ substrate. Furthermore, to demonstrate the performance as an adsorbent in practical applications, we also studied the room temperature stability and selectivity of $Ti_2CN_2$. Our results give theoretical guidance for the improvement of Sr ion adsorption properties by functional group modification of MXene.

## 2. Methodology

All of our calculations were performed by using the Vienna Ab-initio Simulation Package (VASP)[31, 32]. The electron-ion interactions were characterized by the projector-augmented wave (PAW) method[33, 34]. The exchange–correlation functional was described using the Perdew–Burke–Ernzerhof (PBE) version of the generalized gradient approximation (GGA)[35]. The Brillouin zone is sampled on a Γ-centered Monkhorst-Pack $k$-mesh (3×3×1 for structural relaxation and 5×5×1 for electronic structure calculations) with the structure optimized until the convergence criterion of an energy difference of $10^{-5}$ eV and a Hellman-Feynman force of 0.02 eV Å$^{-1}$ on each atom[36]. A cutoff energy of 500 eV is used for the plane wave expansion of the valence electron wave function, and a vacuum layer larger than 20 Å is created in the z-direction to avoid interaction between periodic repetitive images. The van der Waals interaction in this calculation is described by an empirical correction in the Grimme (D3) scheme[37]. Ab initio molecular dynamics simulations in this study were performed in the NVT ensemble, with a step of 1 fs using the Nose–Hoover thermostat for a total timescale of 5~10 ps at 300 K. For better understanding the interaction between Sr and Ti$_2$C MXene, the Heyd–Scuseria–Ernzerhof (HSE06) hybrid function is applied in the density of states calculation[38].

In order to assess the stability of N terminated Ti$_2$C MXene, the formation energy ($E_f$) of the N termination group at each site can be calculated as[39]

$$E_\text{f} = (E_{\text{Ti}_2\text{C}+m\text{N}} - E_{\text{Ti}_2\text{C}} - mE_\text{N})/m \tag{1}$$

where $E_{\text{Ti}_2\text{C}+m\text{N}}$, $E_{\text{Ti}_2\text{C}}$ and $E_\text{N}$ are the energy of Ti$_2$C with $m$ N termination groups, the pristine Ti$_2$C without termination group, and a single N atom in the N$_2$ gas phase, respectively.

To evaluate the stability of metal atom M adsorption on the Ti$_2$CT$_2$ (T=N, O, F or OH) surface, the adsorption energy was calculated using the expression[40]

$$E_\text{ads} = (E_{\text{Ti}_2\text{CT}_2+n\text{M}} - E_{\text{Ti}_2\text{CT}_2} - nE_\text{M})/n, \tag{2}$$

where $E_{\text{Ti}_2\text{CT}_2+n\text{M}}$ is the total energy of the metallized Ti$_2$CN$_2$ monolayer and $n$ is the number of effectively adsorbed metal atoms. $E_{Ti_2CT_2}$ and $E_M$ represent the energy of pristine MXene nanosheet and single M atom in bulk metal, respectively.

## 3. Results and Discussion

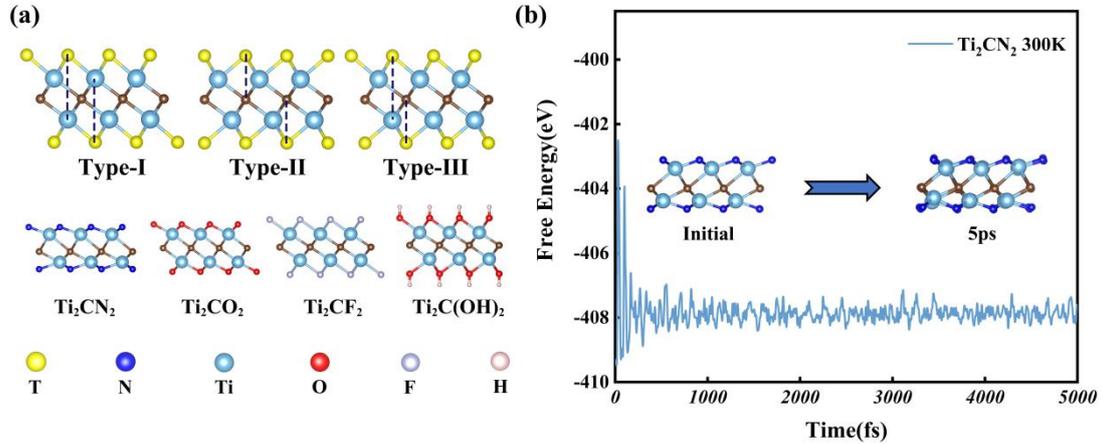

Fig.1. (a) Three possible termination configurations of functional groups on Ti$_2$C surfaces and side views of the most stable configurations of Ti$_2$CN$_2$, Ti$_2$CO$_2$, Ti$_2$CF$_2$ and Ti$_2$C(OH)$_2$; (b) AIMD simulation of Ti$_2$CN$_2$ at 300K.

According to previous study, there are three possible configurations of surface functional group T, corresponding to type I, II and III in Fig.1(a)[41]. In type I all the functional groups are located in the fcc site; in type II, all the surface groups are in the hcp site; and type III is a combination of the above two configurations, with the functional group on one side of it in the hcp site and the other side in the fcc site. In order to obtain the most stable conformations of N, O, F and OH functionalized Ti$_2$C, the energies of all their conformations were calculated as shown in Table S1. The most stable structures (The lowest energy structure) of Ti$_2$CO$_2$, Ti$_2$CF$_2$ and Ti$_2$C(OH)$_2$ were observed in configuration I, while Ti$_2$CN$_2$ preferred configuration III, which is consistent with previous research[42-44]. We find that the terminal functional group type of Ti$_2$C can significantly affect its structural characteristics. The obtained lattice constants for Ti$_2$CN$_2$, Ti$_2$CO$_2$, Ti$_2$CF$_2$ and Ti$_2$C(OH)$_2$ were 3.210, 3.018, 3.036 and 3.052 Å, and their corresponding thicknesses were 3.776, 4.439, 4.795 and 6.791 Å, respectively.

Moreover, the formation energy of 100% N-functionalized Ti$_2$C is −1.739 eV,

which indicates that the formation of $Ti_2CN_2$ structures in the bare-leakage state of $Ti_2C$ under N atmosphere is energetically feasible. The thermodynamic stability of the $Ti_2CN_2$ monolayer was verified by AIMD simulations. As shown in Fig. 1(b), the free energy of $Ti_2CN_2$ monolayer gradually stabilized around –407.87 eV at 300 K, and the structure of $Ti_2CN_2$ without substantial changes throughout the process, which proves the stability of the structure at room temperature.

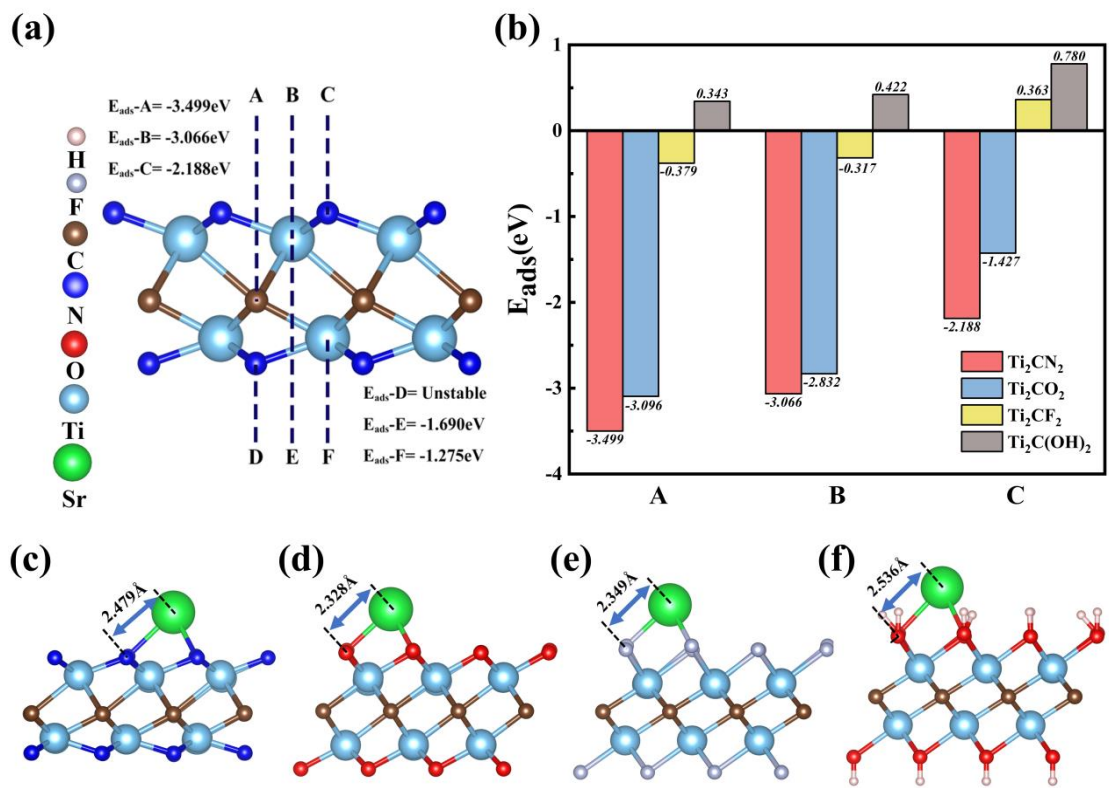

Fig.2. (a) Possible adsorption sites for Sr on $Ti_2CN_2$ MXene; (b) adsorption energy $E_{ads}$ of single Sr ion adsorbed on $Ti_2CT_2$ monolayers at the A, B and C sites; (c)-(f) are the most stable adsorption configurations of individual Sr atom on different surface terminated MXene $Ti_2CT_2$.

Based on the unique structure of $Ti_2CN_2$, there exist six possible adsorption sites on both surfaces. For the upper surface of $Ti_2CN_2$ in Fig. 2(a), the A site is above the inner carbon atom, the B site is above the Ti layer atom, and the C site is above the surface termination group N. For the surface on the other side in Fig. 2(a), the D, E and F sites are above the N, second and first layer Ti atom, respectively. For a comprehensive comparison of the Sr adsorption properties of $Ti_2CN_2$, we calculated the adsorption energy of a single Sr atom on all possible sites of the $Ti_2CT_2$ surface by using equation (2) (*n*=1), as shown in Fig.2(a). Notably, it can be observed that its surface in fcc configuration (A, B and C sites) is more favorable among all considered sites. The adsorption energy at A, B and C sites are responding to –3.499, –3.066 and –2.188 eV, respectively. Actually, for $Ti_2CT_2$ (T=O, F and OH) MXene, there are only three possible adsorption sites A, B and C. In order to facilitate a more lucid comparison with different functional groups, we solely present the adsorption energy at the A, B, and C sites on the surface of $Ti_2CT_2$ (T=N, O, F and OH) as shown in Fig. 2(b). The results show that the individual Sr atom can be effectively adsorbed on the surface of $Ti_2CN_2$, $Ti_2CF_2$ and $Ti_2CO_2$ monolayers. Conversely, the adsorption energies $E_{ads}$ of all three sites on the surface of $Ti_2C(OH)_2$ are positive, which means that it cannot achieve a stabilized capture of Sr ions. Among these three possible adsorption sites, the A site shows the lowest adsorption energy, which means that the A site is the most favorable for Sr atom compared to other positions. In other words, the most stable adsorption of Sr atom occurs at site A. The $E_{ads}$ of sites B and A is close but significantly different from that of site C. This may be attributed to the fact that there are three adjacent termination groups near the A and B sites, which allows Sr atom falling into these sites to form stronger binding with the MXene surface, whereas there is only one at the C site.

Astonishingly, the adsorption energy $E_{ads}$ of a single Sr atom at the most stable site on the $Ti_2CN_2$ surface is –3.499 eV, which is lower than that on $Ti_2CO_2$, $Ti_2CF_2$ and $Ti_2C(OH)_2$, suggesting that the interaction between N and Sr ion is the strongest compared to other termination groups. Through the comparison of the adsorption energies presented in Fig. 2(b), we can determine the stability of Sr ion adsorption by different functional group of $Ti_2C$, which is in the order of: $Ti_2CN_2$ > $Ti_2CO_2$ > $Ti_2CF_2$ > $Ti_2C(OH)_2$. It is striking that the effect of the surface functional group is dramatic, with the adsorption energy $E_{ads}$ of the A site decreasing from 0.780 eV to –3.499 eV after the change of the termination group from OH to N, suggesting that surface termination group plays an important role in the adsorption of Sr ions. Fig.2(c)-(f) show the most stable adsorption configurations of Sr ions on the $Ti_2CT_2$ (T=N, O, F and OH) surface, respectively. Notably, it can be observed that the minimum distance between Sr atoms and surface atoms in any configuration is smaller than the sum of their covalent radii (2.66, 2.61 and 2.52 Å between Sr and N, O and F, respectively)[40], suggesting that the adsorption of Sr atoms on the above substrates may be accompanied by the formation of chemical bonds.

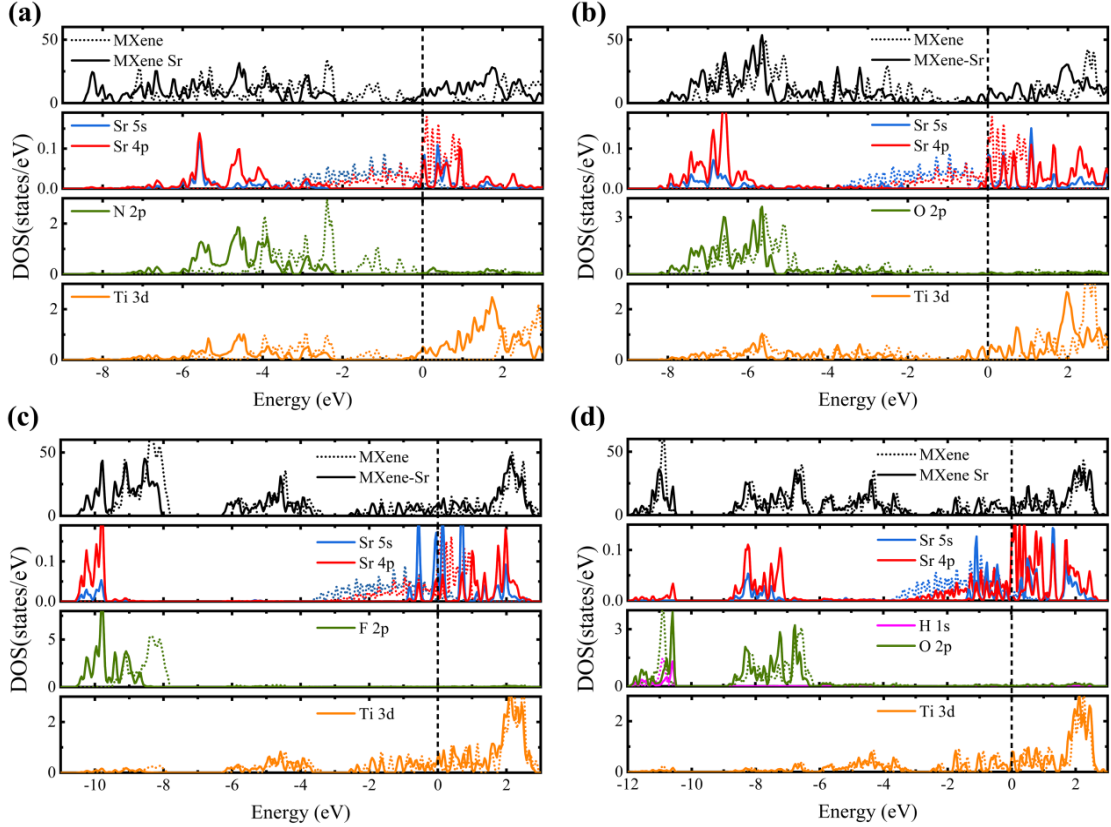

Fig.3. Total density of states (TDOS) and projected density of states (PDOS) of $Ti_2CT_2$-Sr (T= N, O, F, and OH) compounds before and after the adsorption (solid line is corresponding to the case after adsorption, and dotted line is before adsorption).

To reveal the adsorption mechanism of Sr ion on $Ti_2CT_2$ (T=N, O, F and OH), we calculated the total density of states (TDOS) for the most stable adsorption configuration and the projected density of states (PDOS) of the atoms interacting with Sr atom on the substrate before and after adsorption as shown in Fig. 3. Undoubtedly, new peaks of Sr-5s and Sr-4p were observed in all considered adsorption configurations of $Ti_2CT_2$ (T=N, O, F and OH) due to the interaction between Sr and substrate. For $Ti_2CO_2$ and $Ti_2CF_2$, the split Sr-5s and Sr-4p hybridize with O-2p and F-2p in the range of –6.5 to –4.0 eV and around –8.0 eV below the Fermi surface, respectively, which demonstrates the existence of the Sr-O (2.328 Å) and Sr-F (2.349 Å) bonds postulated above. Similarly, in the Ti$_2$C(OH)$_2$-Sr adsorption conformation, it can also be found that the split Sr-5s and Sr-4p hybridize with O-2p in the range of –7.0 to –5.5 eV, with the subsequent formation of Sr-O bond. It is remarkable that the hybridization between H-1s and O-2s is affected after the adsorption of Sr ion. In this process, the O-H bond near Sr ion in Fig.3(c) is elongated and shows a significant offset, which indicates that there is a repulsion interaction between the H atom and the Sr ion. Obviously, although Sr-O bonding is also observed in the Sr-Ti$_2$C(OH)$_2$ system, the intervention of the terminal H atom weakens the Sr-O interaction and results in a significant decrease in the hybridization of Sr-5s and Sr-4p with O-2p, accompanied by the extension of the Ti-O bond to 2.536 Å.

Interestingly, in the case of Ti$_2$CN$_2$, it is evident that N-2p and Ti-3d are hybridized in the –4 to 0 interval, which suggests that the N functional group is firmly bound to the Ti$_2$C surface by means of robust Ti-N bonding. Furthermore, upon Sr ion adsorption, it can be observed that because of the interaction with the substrate, Sr-5s and Sr-4p split new peaks in the region of –6 to –2 eV and hybridize with N-2p to form the N-Sr bond in Fig.2(c).

The charge depletion around Sr ions and the charge accumulation on terminal T can be observed from the charge density difference in Fig. 4, which suggests that the strong interaction exists between Sr ion and Ti$_2$CT$_2$ substrate. A drastic charge transfer occurred after adsorption when the surface termination groups were N, O and F, while the degree of charge transfer was significantly weakened with the replacement of the termination group with OH, which may be attributed to the discrepancy in the electronegativity of the surface groups. Meanwhile, based on the Bader charge analysis, we found that the charge transferred from Sr ion to the substrate is in the range of 0.701~1.561 e. It should be noted that although the charge exchange is mainly concentrated around the functional groups adjacent to Sr, the charge distribution of the whole substrate surface is clearly affected, which indicates that the adsorption properties of Sr ion are not only related to the surface termination groups, but also to the charge properties of the whole substrate (inner layer atomic species).

In addition, the strength of the chemical bond formed between Sr ion and $Ti_2C(OH)_2$ is insufficient to support the effective adsorption of Sr ions due to the repulsive effect between the terminal H atoms of the OH group and Sr ion. In fact, the O-H bonding of the terminal hydroxyl group in $Ti_2C(OH)_2$ is not robust, and the process of capturing Sr ions may be concomitant with the substitution of single or multiple OH-terminated H ions by metal cations[40]. The Sr ions exchange with H ions on the $Ti_2C(OH)_2$ surface are presented in the Supplementary Information.

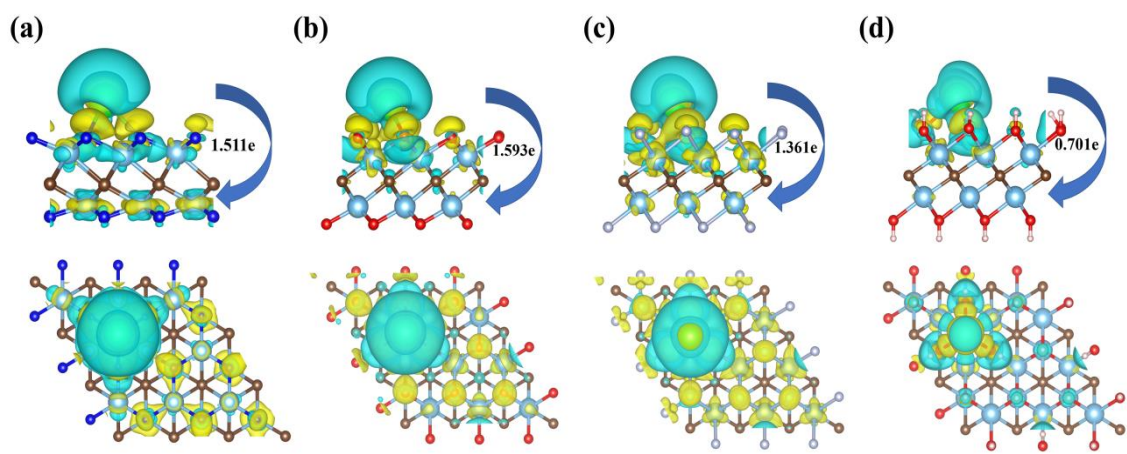

Fig.4. Charge density differences for Sr-$Ti_2CT_2$ (T=N, O, F and OH). Yellow and blue colors with the isosurface value of 0.0015 e/Bohr$^3$ represent charge accumulation and depletion, respectively. The charge density difference is the difference before and after the adsorption of Sr atom on the $Ti_2CT_2$ surface.

The adsorption capacity is an important index to evaluate the performance of the adsorbent. The maximum number ($n_{max}$) of Sr ions for which effective adsorption ($E_{ads}$<0) occurs at monolayer adsorption can be obtained by using Equation S2. The coverage of Sr ions on the MXene surface is defined as $\theta=n/18$ ($n$ is the number of placed Sr ions; 18 is the maximum number of adsorptions in a 3 × 3 cell monolayer). For $Ti_2CN_2$ monolayer, it is shown that the adsorption energy is –0.827 eV for a Sr coverage of 1 ML ($n$=18) on the $Ti_2CN_2$ surface, which indicates that the first layer of Sr ions are all under effective adsorption. Since the configurations of N atoms on both sides of $Ti_2CN_2$ are different, the adsorption energies of 9 Sr ions (1/2 ML) on the upper and lower surfaces of $Ti_2CN_2$ were calculated separately to ensure the reliability of our conclusions. As shown in Fig. 5(a), the $E_{ads}$ obtained in both cases are less than 0 (upper surface –1.043 eV, lower surface –0.107 eV). When Sr ions are placed on the second layer on both sides as shown in Fig. 5(b), the adsorption energies obtained from equation S2 in Supplementary Information are greater than 0 (the positions of Sr in the second layer all point vertically to the N atoms on both sides; upper surface 0.178 eV and lower surface 0.246eV), which may be due to the strong repulsive force between Sr ions hindering the effective adsorption of the second layer. Conversely, Sr ions are unable to reach effective adsorption under full coverage on $Ti_2CO_2$ and $Ti_2CF_2$ surfaces. It can be seen that the maximum coverage of Sr ions on the $Ti_2CO_2$ and $Ti_2CF_2$ surfaces is 8/9 ML and 4/9 ML, corresponding to the maximum adsorption numbers of Sr ions of 16 and 8, respectively. Intuitively, the theoretical adsorption capacities of $Ti_2CF_2$, $Ti_2CO_2$ and $Ti_2CN_2$ monolayers were calculated to be 0.534, 1.115 and 1.291g g$^{-1}$, respectively.

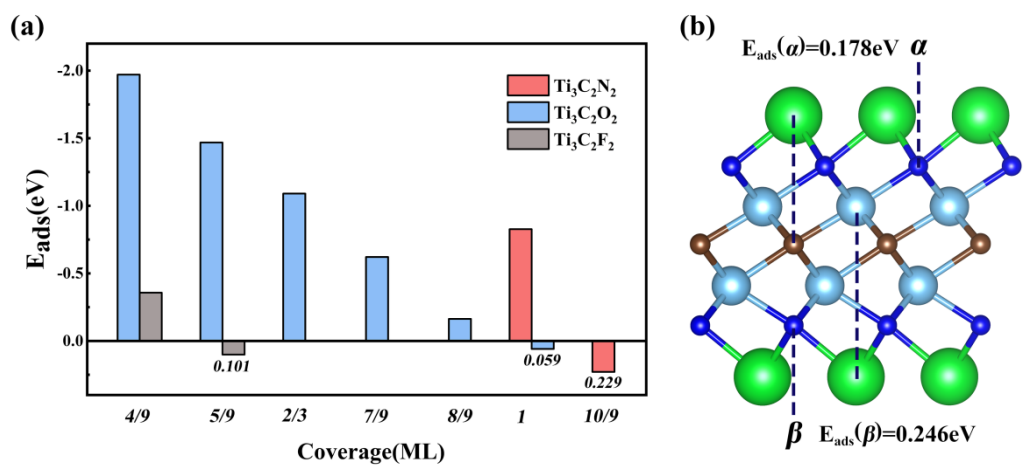

Fig.5. (a) Variation of the adsorption energy of Sr ions on the $Ti_2CT_2$ (T = N, O and F) surface with the coverage. (b) Structure of Sr ions adsorbed on $Ti_2CN_2$. (α and β are the adsorption sites of the second layer of the upper and lower surfaces, respectively)

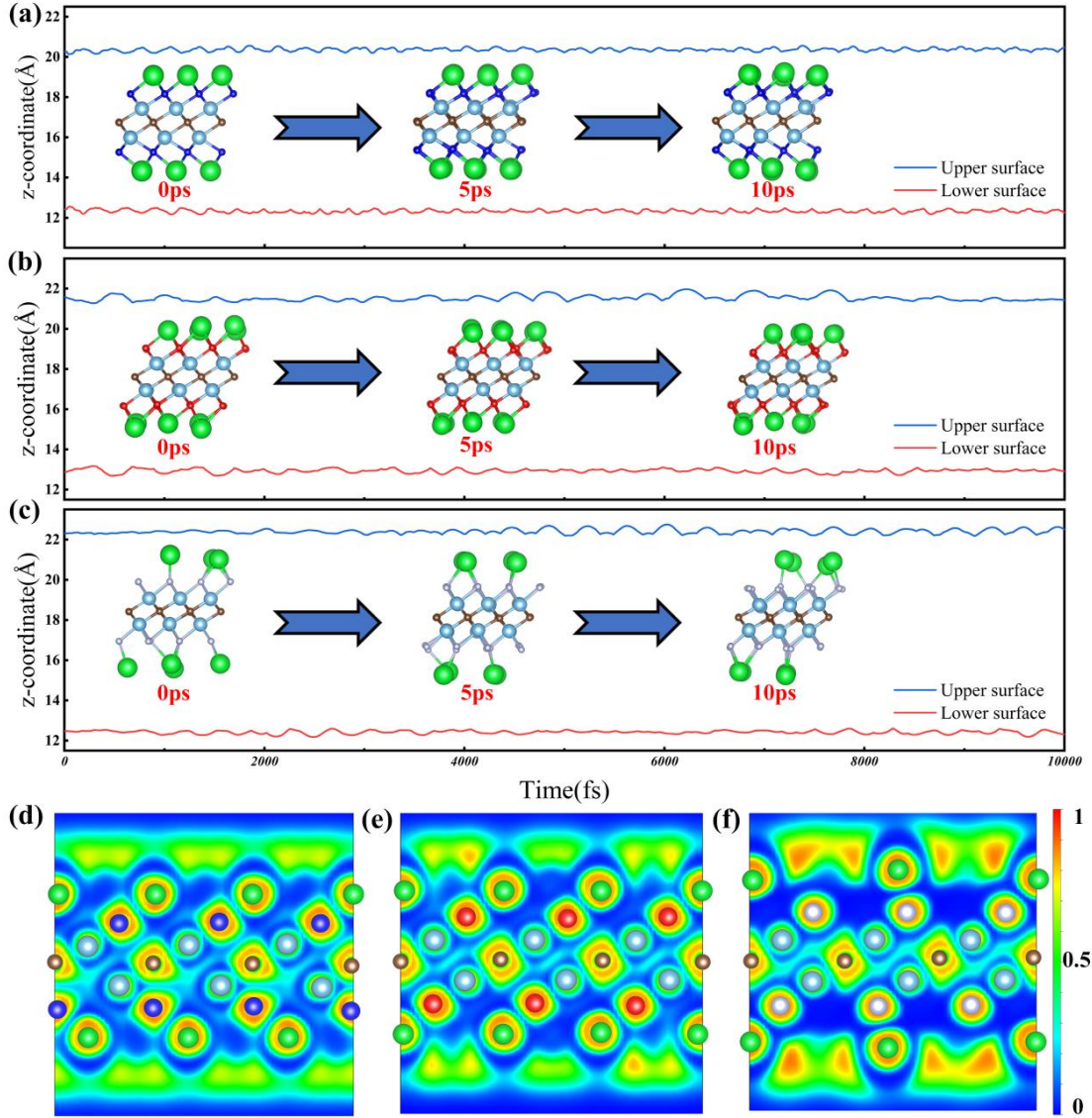

Fig.6. (a)-(c) AIMD simulation of $Ti_2CN_2$, $Ti_2CO_2$ and $Ti_2CF_2$ at 300 K after saturation of Sr ion adsorption (the blue and red lines represent the z-coordinates of the Sr ions adsorbed on the upper and lower surfaces, respectively); (d)-(f) Projections of the electron localization functions of $Ti_2CN_2$, $Ti_2CO_2$ and $Ti_2CF_2$ in the (110) plane under the maximum coverage of Sr ions.

In order to verify the thermal stability of Sr adsorption on $Ti_2CN_2$, $Ti_2CO_2$ and $Ti_2CF_2$ substrate surfaces, AIMD simulations were performed at 300 K with the maximum adsorption of Sr ions. As shown in Fig.6(a-c), the structures of all three substrates did not change essentially after 10 ps, which indicates that their structures are still thermally stable. By tracking the z-coordinate of Sr ions on both sides, it can be observed that Sr ions on the surfaces of $Ti_2CN_2$, $Ti_2CO_2$ and $Ti_2CF_2$ have never shown desorption behavior throughout the 10 ps, which proves that Sr ions can still be stably adsorbed on $Ti_2CN_2$, $Ti_2CO_2$ and $Ti_2CF_2$ substrates at room temperature.

Fig.6(d)-(f) show the electron localization functions (ELF) of the (110) section of $Ti_2CN_2$, $Ti_2CO_2$ and $Ti_2CF_2$ monolayers after reaching the maximum adsorption capacity, respectively. It can be seen that the electrons of Sr ions on $Ti_2CN_2$, $Ti_2CO_2$ and $Ti_2CF_2$ spread out and form a negative electron cloud (NEC), especially the Sr ions on the surface of $Ti_2CN_2$ have been completely surrounded by NEC. When higher concentrations of adsorption occur, the NEC can weaken the mutual repulsive forces between alkaline earth metals (among the positive metal ions)[45]. After reaching the maximum adsorption capacity, the interactions between Sr and $Ti_2CT_2$ (T=N, O and F) substrate can still be observed, which ensures that high concentrations of Sr ions can be stably retained on the substrate surface. Furthermore, the order of the interaction strength ($Ti_2CF_2$ < $Ti_2CO_2$ < $Ti_2CN_2$) is consistent with the maximum number of adsorption, suggesting that the adsorption capacity is closely related to the interaction between the substrate and Sr ions.

Finally, the selectivity of $Ti_2CN_2$, $Ti_2CO_2$ and $Ti_2CF_2$ for Sr ions in the presence of several competing cations was also investigated. The $E_{ads}$ of several common impurity metal ions in water on their surfaces were calculated, as shown in Fig.7. Encouragingly, the adsorption energy of Sr ions on the surface of $Ti_2CN_2$ and $Ti_2CO_2$ is the lowest among Sr, Ca, K, Na, Mg, Al and Zn ions, suggesting that $Ti_2CN_2$ and $Ti_2CO_2$ are able to maintain high selectivity for Sr ions in the presence of Ca, K, Na, Mg, Al and Zn plasmas together. As for $Ti_2CF_2$, it exhibits a higher preference for monovalent cations such as K and Na in comparison. In fact, even though Sr is preferential in energy, the presence of competing cations will inevitably have an impact on its adsorption capacity.

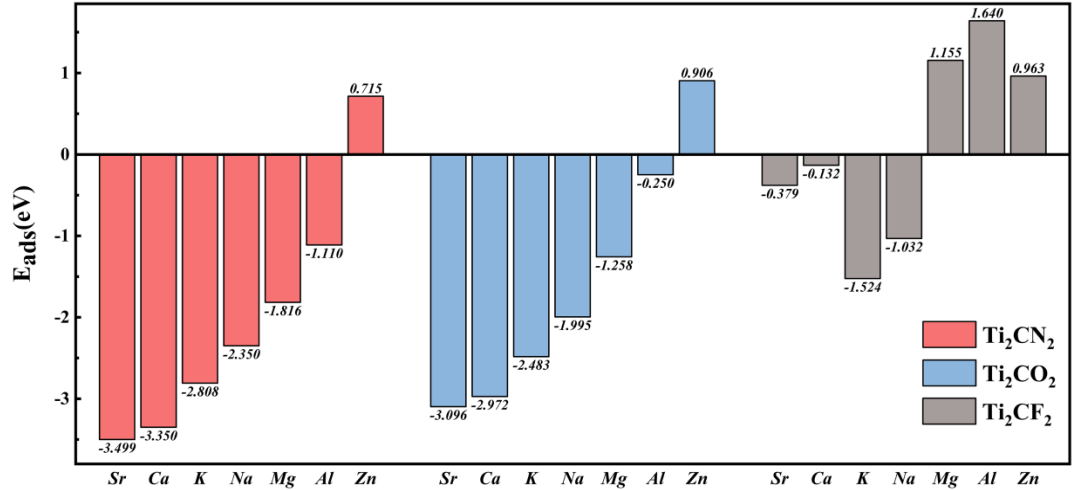

Fig.7. Comparison of the adsorption energies of Sr, Ca, K, Na, Mg, Al and Zn on $Ti_2CN_2$, $Ti_2CO_2$ and $Ti_2CF_2$ surfaces, respectively (all adsorption energies shown are for the most stable sites).

## 4. Conclusion

In this work, we first examined the stability of N-functionalized $Ti_2C$ through first-principles calculations. The results show that the N-functional groups will be connected to the exposed $Ti_2C$ with strong Ti-N bonds and can generate stable structures with completely N-functionalized surface. Subsequently, we systematically investigated and compared the potential of N-functionalized and other functionalized (O, F and OH) $Ti_2C$ as Sr ion adsorbents and dissected their adsorption mechanisms. We found that there is a chemical interaction between the adsorbed Sr ion and these aforementioned functionalized $Ti_2C$, in the order of $Ti_2CN_2 > Ti_2CO_2 > Ti_2CF_2 > Ti_2C(OH)_2$. Regarding the adsorption stability, $Ti_2C(OH)_2$ cannot support the effective capture of Sr ions, while $Ti_2CN_2$, $Ti_2CO_2$ and $Ti_2CF_2$ can all effectively capture Sr ions. Especially the N atoms on the surface of $Ti_2CN_2$ are connected to Sr ions by a robust N-Sr bond, which provides a stronger interaction and greater capacity of Sr ion adsorption on $Ti_2CN_2$ than $Ti_2C$ with other surficial groups. The adsorption capacity of $Ti_2CN_2$ for Sr ions can reach full-layer coverage (1 ML), while the adsorption capacity of $Ti_2CO_2$ and $Ti_2CF_2$ are 8/9 and 4/9 ML, respectively. Furthermore, the thermal stability of Sr adsorption on $Ti_2CT_2$ (T=N, O and F) is verified by using

AIMD at 300 K. We have also studied the selectivity property of $Ti_2CT_2$ (T=N, O and F).The results show that both $Ti_2CN_2$ and $Ti_2CO_2$ exhibit a high Sr selectivity in the presence of multiple competing cations (Ca, K, Na, Mg, Al and Zn), while $Ti_2CF_2$ exhibits a low selectivity for Sr ions in the presence of Na and K ions. This work to be validated in experiments is expected to provide a new perspective for the design of MXenes as adsorbents for nuclides.

**Declaration of Interest Statement**

We declare that no conflict of interest exists.

**Data Availability Statement**

Data are available on request.

**Supporting Information**

The calculated total energy, lattice constants, and thickness of $Ti_2CT_2$ (T=N, O, F, and OH), Sr ions exchange with H ions on the $Ti_2C(OH)_2$ surface, and definition of the adsorption energy of the additional layer.

**Acknowledgment**

This research was supported by the National Natural Science Foundation of China (Grant No.11875004, 11974055 and 22176190), and the Fundamental Research Funds for the Central Universities (Grant No. FRF-GF-20-07B). This work was also supported by USTB Research Center for International People-to-people Exchange in Science, Technology and Civilization (Grant No. 2022KFTS004).